\documentclass[useAMS,usenatbib,a4paper,referee]{mn2e}
\usepackage{graphicx}
\usepackage{amsmath}
\usepackage{txfonts}

\def\aap{A\&A}
\def\apjl{ApJ}

\def\apjs{ApJS}

\title[Diffuse TeV Emission]{Diffuse TeV Emission at the Galactic Centre}
\author[Elizabeth Wommer, Fulvio Melia, and Marco Fatuzzo]{Elizabeth 
Wommer,$^{1}$\thanks{E-mail: ewommer@physics.arizona.edu} Fulvio 
Melia,$^{2}$\thanks{Sir Thomas Lyle Fellow and Miegunyah Fellow 
E-mail: melia@as.arizona.edu} and Marco Fatuzzo$^{3}$\thanks{E-mail: 
fatuzzo@xavier.edu}
\\
\null$^{1}$ Department of Physics, The University of Arizona, Tucson, 
Arizona 85721, USA
\\
\null$^{2}$ Department of Physics and Steward Observatory, The University 
of Arizona, Tucson, Arizona 85721, USA
\\
\null$^{3}$ Physics Department, Xavier University, Cincinnati, OH 45207}

\date{Submitted to MNRAS 2007 December 27}

\begin{document}
\maketitle
\label{firstpage}
\setcounter{figure}{0}

\begin{abstract}
The High-Energy Stereoscopic System (HESS) has detected intense diffuse
TeV emission correlated with the distribution of molecular gas along the
galactic ridge at the centre of our Galaxy. Earlier HESS observations
of this region had already revealed the presence of several point
sources at these energies, one of them (HESS J1745-290) coincident
with the supermassive black hole Sagittarius A*. It is still not entirely
clear what the origin of the TeV emission is, nor even whether it is
due to hadronic or leptonic interactions. It is reasonable to suppose,
however, that at least for the diffuse emission, the tight correlation
of the intensity distribution with the molecular gas indicates a pionic-decay
process involving relativistic protons. In this paper, we explore the possible
source(s) of energetic hadrons at the galactic centre, and their propagation
through a turbulent medium. We conclude that though Sagittarius A* itself 
may be the source of cosmic rays producing the emission in HESS J1745-290,
it cannot be responsible for the diffuse emission farther out. A distribution
of point sources, such as pulsar wind nebulae dispersed along the galactic
plane, similarly do not produce a TeV emission profile consistent with the
HESS map. We conclude that only a relativistic proton distribution accelerated
throughout the inter-cloud medium can account for the TeV emission profile
measured with HESS.
\end{abstract}


\section{Introduction}
Observations of the galactic-centre ridge with the High-Energy Stereoscopic System
(HESS) have revealed a surprisingly intense diffuse TeV emission, strongly correlated
with the distribution of interstellar gas (Aharonian et al. 2006). This feature,
along with the energy range accessible to HESS ($> 200$ GeV), suggests that
the dominant component of this diffuse emission is almost certainly due to the decay
of neutral pions produced in hadronic cascades initiated via the scattering of
relativistic cosmic rays with protons in the ambient medium (see, e.g., Crocker
et al. 2005; Ballantyne et al. 2007).

Earlier HESS observations of the galactic centre had already revealed the
presence of several TeV point sources, one of these (HESS J1745--290) coincident
with the supermassive black hole Sagittarius A* (Aharonian et al. 2004). A second
nearby source (about 1$^\circ$---or roughly 144 pc at that distance---toward positive
longitude $l$ of the galactic centre) was identified with the supernova remnant/pulsar
wind nebula G0.9+0.1. With HESS's unprecedented sensitivity, it has been possible
to subtract these (and other) point sources from the overall map of this region, to
search for the fainter, diffuse emission. The latter extends along the galactic
plane for over 2$^\circ$, and is also spread out roughly 0.2$^\circ$ in galactic
latitude $b$.

At the distance to the galactic centre, an extension in latitude of $\sim 0.2^\circ$
corresponds to a scale height of about 30 pc, similar to that of giant molecular
cloud (GMC) material in this region, as traced by its CO and CS line emission
(see, e.g., Tsuboi et al. 1999). These data suggest that the Galaxy's central region
($|l|<1.5^\circ$ and $|b|<0.25^\circ$) contains up to $\sim 10^8\;M_\odot$ of
molecular gas, providing a rich target of overlapping clouds for the incoming
cosmic rays (see figure~1; the construction of this diagram is discussed in
\S\ 2 below).

\begin{figure}
\center{\includegraphics[scale=0.40,angle=0]{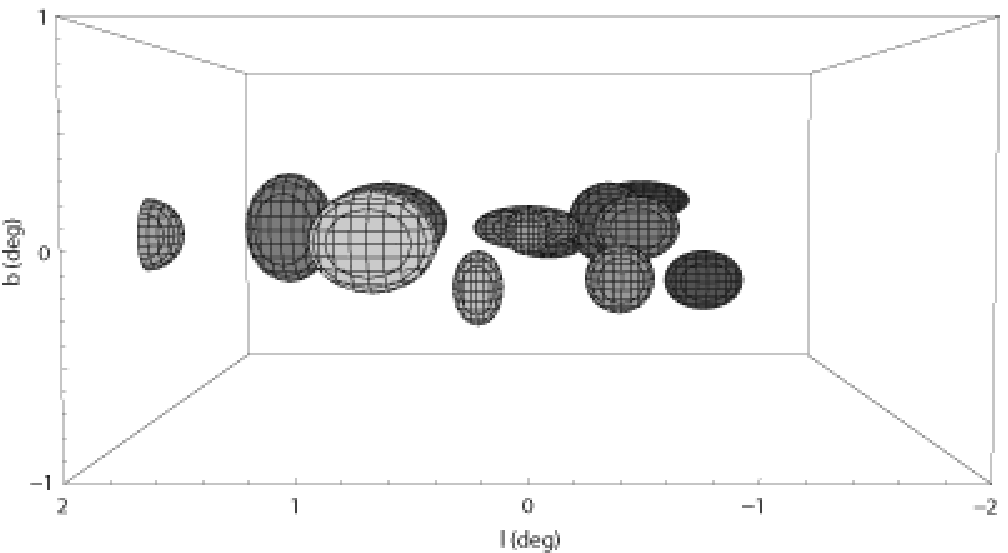}
\hspace{0.3in}\includegraphics[scale=0.40,angle=0]{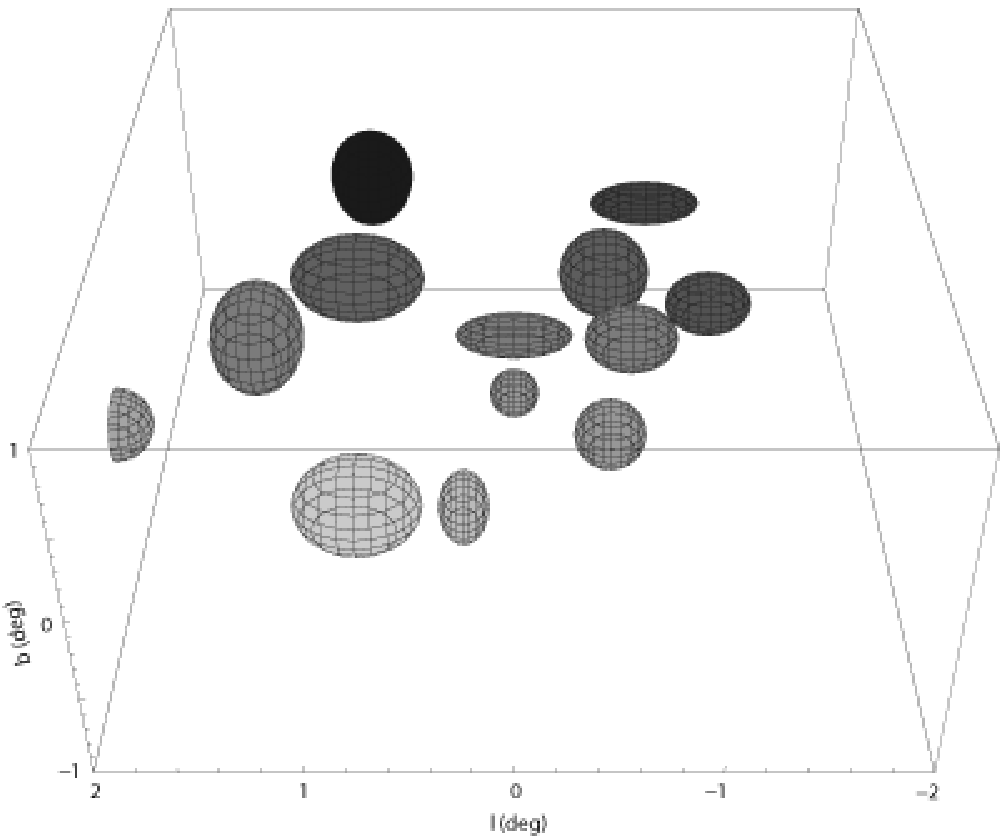}
\vspace{10pt}
\caption{Cloud distribution with the first assignment of $z$-coordinates.
The left panel shows the line-of-sight view of the molecular gas. The right
panel represents the same view, though from a vantage point slightly above the 
galactic plane.}}
\end{figure}

However, a simple cosmic-ray interpretation for the diffuse TeV emission is
problematic for several reasons. Chief among these is the observed gamma-ray
spectrum, which does not appear to be consistent with the cosmic-ray
distribution in the solar neighbourhood. The reconstructed gamma-ray spectrum
for the region $|l|<0.8^\circ$ and $|b|<0.3^\circ$ (with point-source emission
subtracted) is a power law with photon index $\Gamma=2.29\pm 0.27$. Since
for a power-law energy distribution the spectral index of the gamma rays
tracks the spectral index of the cosmic rays themselves, the implied
cosmic ray index ($\sim 2.3$) is much harder than that ($\sim 2.75$)
measured at Earth. Of course, we must be aware of the fact that cosmic
rays escape from the Galaxy as they diffuse outwards from the centre,
on an energy-dependent time scale $t_{\rm esc}\propto E^{-\delta}$, with
$\delta\sim0.4$--$0.6$ (Bhattacharjee 2000). Thus, if most of the cosmic
rays detected at Earth indeed originate at the galactic centre, then the
injected spectrum ought to be flatter (by a change in index of $\sim\delta$)
than that observed here, and the two distributions may still be consistent with
each other. This very interesting possibility deserves further analysis,
which may be facilitated by the results presented in this paper.

Secondly, the diffuse spectrum is remarkably similar to that of the central
point source HESS J1745--290 itself ($2.29\pm 0.27$ versus $2.25\pm 0.10$).
Though other sources (such as pulsar wind nebulae) along the galactic ridge
also have similar spectra, the central TeV source is by far the brightest
and may therefore be the dominant hadron accelerator in this region. It is
indeed of considerable interest to know whether a single source (say, the
supermassive black hole) could be responsible for producing most of the
relativistic particles in the Galaxy's central region, regardless
of whether or not these should rightly be regarded as members of the
overall cosmic-ray population.

In an earlier paper (Ballantyne et al. 2007), we examined the possible
role of Sagittarius A* in producing the central point source HESS J1745--290,
and concluded that stochastic acceleration within the inner 20--30 Schwarzschild
radii of the black hole's event horizon could produce both the relativistic
electrons responsible for Sagittarius A*'s mm-spectrum, as well as an
outflowing flux of relativistic protons that diffuse outwards to fill the
inner 2--3 pc region, where they scatter predominantly with molecular gas
in the circumnuclear disk to produce the TeV signal. Only $\sim 1/3$ of these
protons encounter the disk, however, so up to this point, it is still not
known whether the remainder of these particles can diffuse outwards to much
larger radii to produce the diffuse emission along the galactic ridge.

In addition, it is not yet clear whether Sagittarius A* itself is the
main contributor of relativistic hadrons to the source HESS J1745--290.
At least two alternative models have been proposed, including a plerion
scenario with Sagittarius A* as the wind source (Atoyan and Dermer 2004),
and a pulsar wind nebula discovered recently within only a light-year
of Sagittarius A* (Wang, Lu, and Gotthelf 2005). But all leptonic
models for the TeV emission in HESS J1745--290 (and possibly elsewhere
along the ridge) are subject to the extremely rapid cooling rates resulting
from the intense photon fields and relatively strong magnetic fields in and
around the molecular clouds in this part of the Galaxy.

One of the principal goals of this paper is therefore to examine whether
relativistic protons accelerated by Sagittarius A* can in fact fill the
$\sim 2^\circ$ region surrounding the black hole to account for the
diffuse TeV emission as well. We will conclude that this scenario is
very unlikely. However, adhering to the idea that hadronic cascades,
rather than inverse-Compton-scattering lepton distributions, are more
likely to produce the diffuse TeV signal, we will also consider other
source configurations that produce a TeV emissivity consistent with
the HESS data.

In subsequent sections, we will first assemble the data pertaining to
the molecular gas distribution at the galactic centre, and then
describe our approach in setting up a turbulent magnetic field
through which the hadrons must diffuse. As they wind their way
outwards from their respective source(s), the protons lose energy
steadily before finally scattering with other protons, and we
will examine the processes that dominate this energy loss rate.
We will describe our technique for calculating the proton
propagation through this medium, which does not rely on a
``standard" diffusion approach, but is instead more robust and
less dependent on unknown factors, such as the diffusion
coefficients. Finally, we will summarize our method for
calculating the particle cascade (once a collision has occurred),
and describe our simulations and results.

\section{The Molecular Cloud Distribution}
Observations of the inner few hundred parsecs of the Galaxy reveal 
a large concentration (up to $\sim 10^8\;M_\odot$) of dense molecular
gas (G\"{u}sten and Philipp 2004). Much of this material is concentrated
within giant molecular clouds, with a size $\sim 50$--$70$ pc. Compared 
with their counterparts in the galactic disk, the galactic centre (GC) 
molecular clouds are denser and warmer. Emission maps of density-sensitive 
molecular species, e.g., CS (Bally et al. 1985), H$_2$CO (G\"{u}sten and 
Henkel 1983), and HC$_3$N (Walmsley et al. 1986), indicate that the
GC clouds are ``clumpy", with high-density ($\sim 10^5$ cm$^{-3}$) 
regions embedded within a less dense ($\sim 10^{3.7}$ cm$^{-3}$) intra-cloud 
medium. The average cloud density is then roughly $10^4$ cm$^{-3}$, which 
is quite large compared with the value $\sim 10^{2.5}$ cm$^{-3}$ for a 
typical disk cloud, but necessary if the GC clouds are to survive the 
strong tidal forces in the galactic-centre potential.

The temperature of GC clouds is also relatively high ($\sim 30$--$60$ K) 
and fairly uniform (Morris et al. 1983). Cloud temperatures were first
estimated using measurements of metastable transitions of ammonia 
(NH$_3$) and later confirmed by AST/RO ({\bf A}ntarctic {\bf
S}ubmillimeter {\bf T}elescope and {\bf R}emote {\bf O}bservatory) 
observations of CO line emission (see, e.g., Kim et al. 2002).
Diffuse X-ray emission detected from the galactic centre by {\it Chandra} 
indicates the presence of ``soft" ($0.8$ keV) and ``hard" (8 keV) plasma
components coexisting with the giant molecular clouds (Muno et al. 
2004). While supernova shock waves can provide an explanation for
the cooler plasma, the origin of the hard component is still uncertain.  
Clearly, though, the galactic centre is a warm environment and many 
possible heating mechanisms have been proposed to explain the observed 
cloud temperatures. Direct heating by energy dissipation via collisions 
with dust has mostly been discounted due to the comparatively low 
temperature of the dust particles ($21\pm 2$ K; Pierce-Price et al. 
2000). Other possibilities include magnetic viscous heating, small 
scale dissipation of supersonic turbulence, large scale J- and C-shocks, 
and UV-heating in exposed photo-dissociation layers. More than likely, 
all of these mechanisms contribute on some level, but the dissipation 
of supersonic turbulence appears to be the most promising means by 
which the majority of the molecular clouds are warmed as the heating 
rate of this process is comparable to the cooling rate of the gas.

The GC molecular clouds are also threaded by a pervasive magnetic
field, whose strength is revealed, e.g., by the presence of 
non-thermal filaments (NTFs) in the diffuse interstellar medium (ISM). 
The NTFs appear in radio images as long thin strands (tens of pc in 
length while only fractions of a pc wide) more or less perpendicular 
to the plane of the galaxy (Morris 2007). The strongly polarized 
synchrotron emission from the NTFs indicates that the magnetic field 
points along the filaments, whose apparent rigidity when they 
interact with molecular clouds and the turbulent interstellar medium 
suggests field strengths on the order of a few milligauss (see, e.g., 
Yusef-Zadeh and Morris 1987). Whether the NTFs are manifestations of 
a large-scale poloidal magnetic field or are localized structures 
has not been firmly established.

Mid- and far-infrared thermal dust emission from magnetically aligned 
dust grains provides additional information about the direction of 
the magnetic field within the warm, dense molecular clouds. In direct 
contrast to the ISM field, the magnetic field within the clouds is,
for the most part, parallel to the galactic plane (Werner et al. 1988).  
Clues regarding how the galactic centre can have both poloidal and 
toroidal magnetic field configurations are provided by submillimeter 
polarization measurements of the central molecular zone by Chuss et al. 
(2003), which show that the field orientation is linked to the density 
of the region. Differential motion between dense material and the 
surrounding medium shears the poloidal magnetic field making it more 
toroidal, whereas less dense regions (like the cloud envelope) do not 
have sufficient mass to distort the initial field.                 

The cause of the stability of the GC giant molecular clouds has also 
been the subject of much discussion. It had been shown that, in general, 
the CO luminosity of a molecular cloud scales with its virial mass 
(Young and Scoville 1991). Using this result, the molecular mass of a 
region could be determined from its measured CO luminosity by using a 
standard conversion factor $X \equiv N(H_2)/I_{CO} = 3.0\times 10^{20} 
\;({\rm cm}^{-2}\; [{\rm K}\;{\rm km}\;{\rm s}^{-1}]^{-1})$. However, 
measurements by Oka et al. (1998) show that GC molecular clouds do 
not follow the $L_{CO}$--$M_{VT}$ trend of galactic disk clouds. 
Simply changing the $X$-factor used for GC clouds is not an option 
since its value is set by observations of $\gamma$-ray (Blitz et al. 
1985) and far-infrared (Cox and Laureijs 1989) emission. Based on 
these findings, Oka et al. (1998) conclude that GC clouds larger 
than $\sim 30$ pc are not bound by their self-gravity but are instead
in equilibrium with the external pressure of the galactic centre 
environment. However, the observed pressure due to a hot plasma $P_{plasma} 
\sim 10^{-9.2} \;{\rm erg}\;{\rm cm}^{-3}$ is an order of magnitude 
smaller than that required since the turbulent pressure within the clouds is 
$P_{turb} \sim 10^{-8} \;{\rm erg}\;{\rm cm}^{-3}$ (G\"{u}sten 
and Philipp 2004). Clouds may instead be bound by their own magnetic fields. 
Equating the turbulent and magnetic ($B^2/8\pi$) energy densities
gives field strengths of $\sim 0.5$ mG within the clouds. Then, using 
the pressure-bound assumption, Oka et al. (1998) infer the value
$2 \times 10^7 \;M_\odot$ for the lower limit to the mass of molecular 
gas within the inner 400 pc of the galaxy.

In spite of this abundant material, however, it is somewhat surprising 
that the star formation rate (SFR) in the galactic centre is currently 
not especially high ($\sim 0.3$--$0.6\;M_\odot$ yr$^{-1}$ in the central 
500 pc as opposed to $\sim 5.5\;M_\odot$ yr$^{-1}$ in the disk;  G\"{u}sten 
2004, and references cited therein). Detection of 6.7 keV line-emission 
due to a helium-like iron K-shell transition by Yamauchi et al. (1990), 
though, suggests that either $\sim$1,000 supernova explosions, or one 
$10^{54}$-erg explosion, occurred in this region within the past $10^5$ 
years. Their results lend independent support to the earlier work of 
von Ballmoos, Diehl, and Sch\"{o}nfelder (1987), who investigated the 
1.8 MeV $\gamma$-ray emission from $^{26}$Al, produced in nuclear 
reactions during energetic events like supernova explosions. These
observations indicate that the galactic centre may have been a more 
active star-forming region in the recent past.

\begin{figure}
\center{\includegraphics[scale=0.40,angle=0]{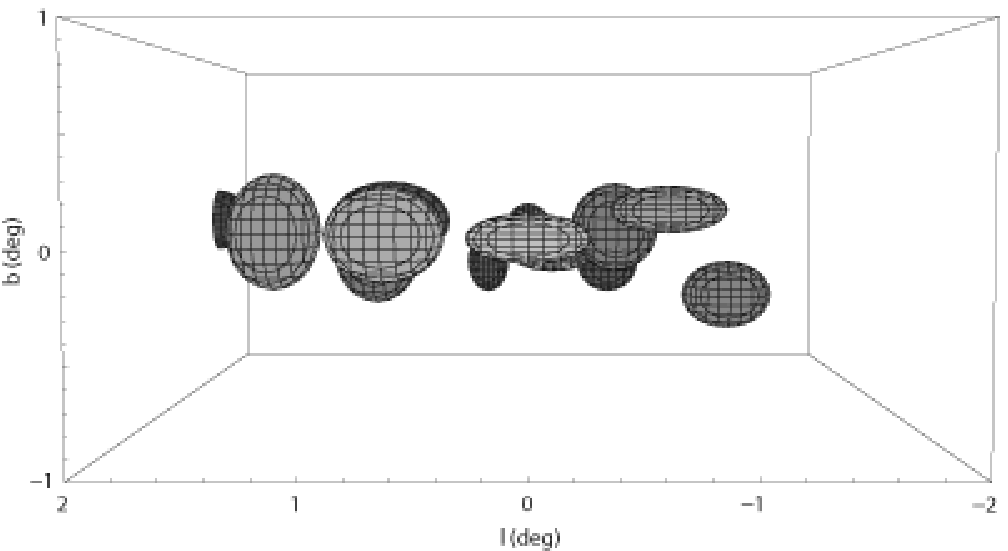}
\hspace{0.3in}\includegraphics[scale=0.40,angle=0]{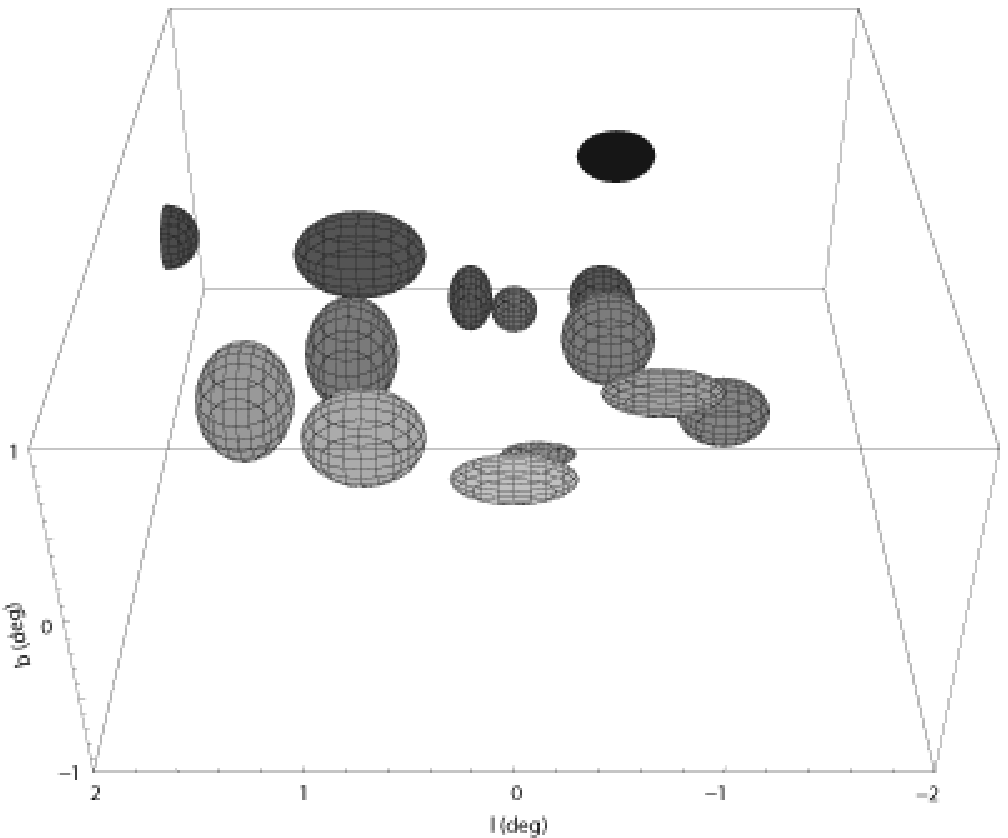}
\vspace{10pt}
\caption{Same as figure~1, except for the second (different) assignment
of $z$-coordinates.}}
\end{figure}

Other than the physical conditions within these clouds, the remaining
ingredient required to assemble a reasonable representation of the molecular
gas distribution along the galactic ridge is the three-dimensional spatial
positioning of the cloud centroids. The positions in the plane of the sky 
of the 14 dominant giant molecular clouds (see Table 1) in the galactic centre 
region $(\vert l \vert\le 2^\circ, \vert b \vert \le 1^\circ)$ were taken from 
Oka et al. (1998). For convenience, we divided the longitudinal coordinate 
into 10 bins, each with a width $\Delta l = 0.4^\circ$, and tabulated the 
number of clouds whose centre occurs in the range between $l_i$ and 
$l_i + \Delta l$. The cloud spatial distribution along the line-of-sight 
(i.e., the $z$-direction) is not known, but we used a randomization 
procedure to attribute a value of this coordinate to each cloud, constrained 
by the requirement that the distribution in $z$ matches the distribution 
in $l$. For each cloud in bin $i$ in the longitudinal direction, we assigned 
a randomly chosen cloud to the corresponding bin along the line-of-sight 
direction. Then, each cloud in bin $i$ was given a line-of-sight position 
$z = z_i + \chi\Delta z$, where $z_i$ is the lower bound on bin $i$, $\chi$ is a 
randomly chosen number between 0 and 1, and $\Delta z = \Delta l = 0.4^\circ$ 
is the bin size. Clouds were re-binned and/or re-positioned if overlap between 
clouds occurred. Using this method, we produced two different molecular cloud 
maps, shown in figures~1 and 2.

\begin{table}
\caption{Positions and Sizes of Giant Molecular Clouds at the galactic Centre}
\footnotesize
\begin{tabular}{lllllll}

 Cloud\qquad\qquad&  $l$\qquad\qquad\qquad&  $b$\qquad\qquad\qquad&  
$z_1$\qquad\qquad\qquad&  $z_2$\qquad\qquad\qquad&  $\Delta l$\qquad\qquad\qquad&  $\Delta b$\\
 &  (deg)&  (deg)&  (deg)&  (deg)&  (deg)&  (deg)\\
\hline
 1&  -1.125&  -0.375&  0.97&  -0.04&  0.5&  0.375\\
 2&  -0.75&  0.125&  1.27&  -0.51&  0.625&  0.25\\
 3&  -0.625&  0&  0.12&  2.05&  0.5&  0.375\\
 4&  -0.5&  -0.25&  -0.38&  0.85&  0.375&  0.375\\
 5&  -0.5&  0&  0.74&  0.12&  0.5&  0.5\\
 6&  -0.125&  -0.125&  0.31&  -0.69&  0.375&  0.125\\
 7&  0&  0&  -0.35&  0.39&  0.25&  0.25\\
 8&  0&  0&  0.15&  -1.03&  0.625&  0.25\\
 9&  0.25&  -0.25&  -0.95&  0.85&  0.25&  0.375\\
 10&  0.75&  0&  -1.22&  -0.71&  0.625&  0.5\\
 11&  0.875&  -0.125&  1.99&  0.15&  0.5&  0.625\\
 12&  0.875&  0&  0.69&  0.92&  0.75&  0.5\\
 13&  1.375&  0&  0.13&  -0.42&  0.5&  0.625\\
 14&  2&  0&  -0.61&  1.1&  0.375&  0.375\\
\end{tabular}
\end{table}

\section{The Turbulent Magnetic Field}
After leaving the acceleration zone, protons random-walk their way outwards scattering
off of the relatively strong turbulent magnetic field. Our treatment here differs from
the usual diffusion approach, which is often limited by poorly known factors, such as the
diffusion coefficients. Instead, we follow the motion of individual particles by
solving the Lorentz force equation with a magnetic field whose spatial profile 
is consistent with Kolmogorov turbulence. To create this magnetic field, we adopt 
the prescription of Giacalone and Jokipii (1994). While their original aim was to use 
the simulated field to calculate the Fokker-Planck coefficients based on the motion of 
individual particles and then compare those values to analytic theory, we will simply 
use their algorithm to track the particle trajectories themselves.

For a particle of mass $m$ and charge $q$ moving in a magnetic field ${\bf B}({\bf r})$, 
we define ${\bf\Omega}({\bf r}) = q{\bf B}({\bf r})/mc$, where $c$ is the speed of light. 
The total field ${\bf\Omega}({\bf r})$ is then written as the sum of two terms: 
${\bf\Omega}_0$ corresponding to the background field and $\delta{\bf\Omega}$, which is 
the fluctuation about the mean and is not necessarily small. A three-dimensional field 
is then obtained by summing over a number of randomly polarized transverse waves. 
A form for $\delta{\bf\Omega}$ that satisfies Gauss's law $\nabla\cdot{\bf B}=0$ is
\begin{equation}
\delta{\bf\Omega}({\bf r}) = \sum_k \Omega(k)[\cos\alpha(k)\hat{\bf y'} \pm i 
\sin\alpha(k)\hat{\bf z'}] \exp[ikx' + i\beta(k)]\;,
\end{equation}
where $\alpha(k)$ and $\beta(k)$ are random numbers between 0 and $2\pi$. 
The primed and unprimed coordinates are related by the rotation matrix
\begin{equation}
{\bf r'} =
        \left(
        \begin{array}{ccc}
          \cos\theta \; \cos\phi &\cos\theta \; \sin\phi  &\sin\theta \\
          -\sin\phi  &\cos\phi  &0 \\
          -\sin\theta \; \cos\phi  &-\sin\theta \; \sin\phi  &\cos\theta \\
        \end{array}
        \right) {\bf r}
\;.
\end{equation}
The angles $\theta$ and $\phi$ are functions of $k$, such that $0 \le \theta(k) 
\le \pi$ and $0 \le \phi(k) \le 2\pi$. Thus, to create a three-dimensional field, 
five random numbers $(\alpha, \beta, \theta, \phi, \pm)$ are needed for each value 
of k.

Assuming the irregular field is generated by Kolmogorov turbulence, we set
\begin{equation}
\Omega(k) = \Omega(k_{min}) (k/k_{min})^{-\Gamma/2}\;,
\end{equation}
where $k_{min}$ is the wavenumber corresponding to the longest wavelength and 
$\Gamma = 5/3$ is the power of the Kolmogorov spectrum. Although we have
chosen this specific value of $\Gamma$, we do not expect that a particular 
choice of turbulence significantly affects the results of the simulation.

The value of $\Omega(k_{min})$ can be found using the total energy density
\begin{equation}
S = \sum_k \frac{B^2(k)}{8\pi} = \frac{m^2 c^2}{8\pi q^2} \Omega^2(k_{min}) 
\sum_k (k/k_{min})^{-\Gamma/2}\;,
\end{equation}
taking $S = B_0^2/{8\pi}$, so that there is as much energy density contained in 
the fluctuations as there is in the background field. For our simulations, we 
use 200 values of k evenly spaced on a logarithmic scale with wavelengths between 
$0.1$ and $10r_{gyr}$ where $r_{gyr}$ is the gyroradius of a proton in the 
background magnetic field $B_0$.

There are two approaches to implementing a turbulent magnetic field generated 
by this method. The first approach (and the one used by Giacalone and Jokipii 1994) 
is to calculate the magnetic field at every time step for each particle position. 
For a particle moving with relativistic velocity {\bf u}~$\equiv \gamma${\bf v}
(in terms of the velocity vector {\bf v}), its position is found by solving the 
equation of motion
\begin{equation}
\frac{d{\bf u}}{dt} = \frac{1}{\gamma} ~{\bf u} \times {\bf\Omega}({\bf r})\;.
\end{equation}
For each time step, the same set of random numbers is used for each particle so that 
only ${\bf r}$ changes. In the second approach, the magnetic field is generated for a 
given volume at the beginning of the simulation. Giacalone and Jokipii (1994) estimate 
that the required volume is $V \sim 100\lambda_{max}^3$, where $\lambda_{max}$ is the 
longest wavelength of the fluctuations. Taking the interpolation distance between
volume lattice points to be $0.25 \lambda_{min}$ ($\lambda_{min}$ being the shortest 
wavelength) results in $6,400(\lambda_{max}/\lambda_{min})^3$ calculations. Since 
we have $\lambda_{max} = 100 \lambda_{min}$, $6.4 \times 10^9$ computations
would have to be performed before any simulations of particle motion can begin. 
This is not only time-consuming, but also very memory-intensive. So, like Giacalone 
and Jokipii (1994), we adopt the former approach. In this way, the magnetic field 
is generated only where needed, the overwhelming amount of computer memory 
required by the second approach is eliminated, and the particles are not confined
to any pre-assigned volume.

\section{Energy Loss Rates}
The protons lose energy through various interactions with the environment as
they diffuse. The dominant energy loss mechanisms are as follows:

\begin{enumerate}

\item{$pp$ scattering}

Defining the energy loss rate as
\begin{equation}
R\equiv -{1\over E}\left({dE\over dt}\right)\;,
\end{equation}
we have for relativistic protons cooling due to their inelastic collisions 
with ambient protons of density $n_p$ 
\begin{equation}
  R_{pp} = n_p c \sigma_{pp} K_{pp} \;.
\end{equation}
The cross section $\sigma_{pp}$ depends only weakly on proton energy, increasing 
from $\simeq$ 30 mbarn at $E_{proton} \sim$ few GeV to 40 mbarn at $10^3$--$10^4$ 
GeV (Karol 1988), so for simplicity we approximate it with a constant value 
$\sigma_{pp}$ = 40 mbarn.  Likewise, although the inelasticity parameter $K_{pp}$ 
depends on the centre of momentum energy $\sqrt{s}$ (i.e., $K_{pp} = 1.35 s^{-0.12}$ 
for $\sqrt{s} \ge$ 62 GeV, and $K_{pp} = $ 0.5 for $\sqrt{s} \le$ 62 GeV; see 
Markoff et al. 1997, 1999), we use only the low-energy $K_{pp} = $ 0.5 in our calculations.

\item{$p\gamma$ scattering}

An inelastic scattering between a proton and photon may lead to pair production, 
$p \gamma \rightarrow p e^+ e^-$, and photo-pion production, $p \gamma \rightarrow p 
\pi^0$ and $p \gamma \rightarrow n \pi^+$.  The cross-section and inelasticity for 
these processes depend on the photon energy (see, e.g., Begelman, Rudak, and Sikora 
1990). In the proton rest frame, the threshold photon energy for pair production is 
$E^{'(e)}_{th} = 2 m_e \sim 1$ MeV and for pion production is $E^{'(\pi)}_{th} = 
m_{\pi} \left( 1 + m_{\pi}/2m_p \right) \sim 145$ MeV.

For the conditions of interest to us here, it can be shown easily by way of estimate 
that the energy-loss rate due to $p\gamma$ interactions is insignificant compared to 
that from $pp$ scatterings, so we may safely ignore this process here.

\item Synchrotron Processes

The synchrotron cooling rate is

\begin{equation}
  R_{synch} = \frac{4}{3} \left({\frac{m_e}{m_p}}\right)^3 \frac{c \sigma_T u_B}{m_e c^2} \gamma_p \,,
\end{equation}
where $u_B = {B^2}/{2 \mu_0}$ is the (total) magnetic field energy density and 
$\sigma_T = 0.665 \times 10^{-28}$ m$^2$ is the Thomson cross-section.

\item Compton scattering

Finally, the cooling rate due to Compton scattering is
\begin{equation}
  R_{C} = \frac{u_{rad}[x < \frac{m_p}{m_e} \gamma_p^{-1}]}{u_B} R_{synch}\;,
\end{equation}
where $u_{rad}$ is the radiation energy density.

\end{enumerate}

\begin{figure}
\center{\includegraphics[width=8cm]{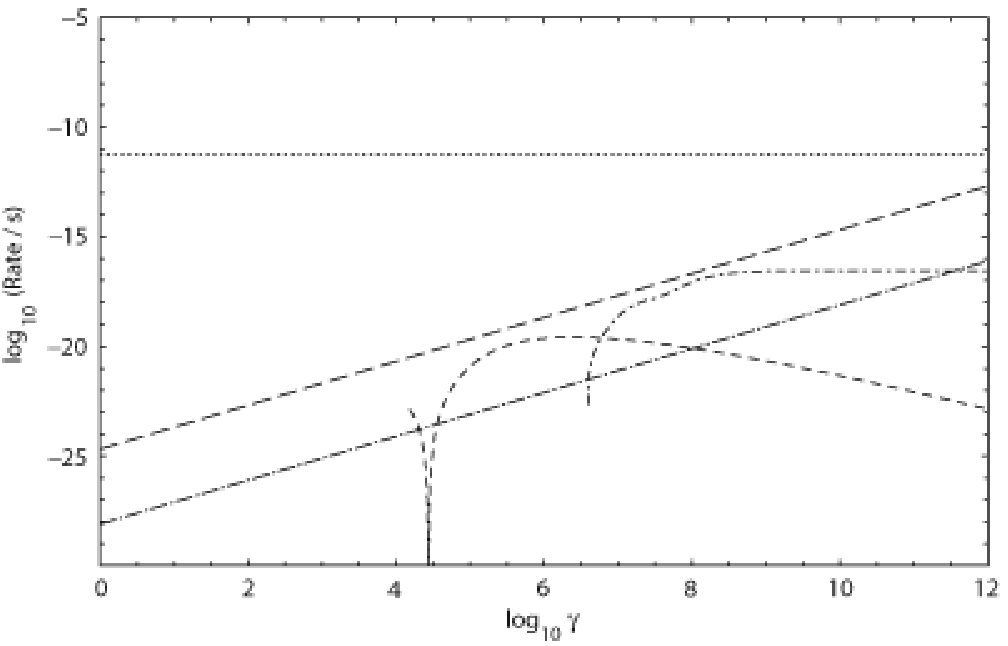}\hskip0.1in
\includegraphics[width=8cm]{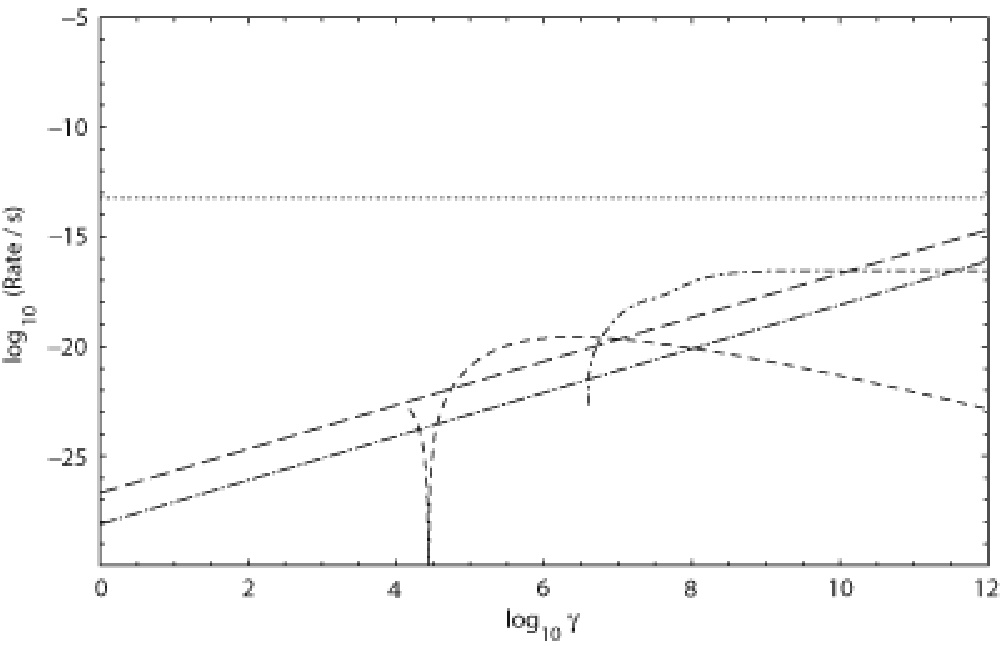}
\vspace{10pt}
\caption{Cooling rates as functions of the proton Lorentz factor 
$\gamma_p$ within the molecular clouds (left) and in the region between the
clouds (right). The dotted line shows the $pp$ scattering cooling rate, 
the long-dashed line is the Compton scattering cooling rate,
the dash-dot line is the synchrotron cooling rate, the short-dashed 
line is the $p \gamma$ pair production cooling rate, and the short
dash-dot line is the $p \gamma$ pion production cooling rate.}
\label{fig:coolingrates}}
\end{figure}

The various cooling rates are plotted in Figure \ref{fig:coolingrates} 
as functions of the proton Lorentz factor $\gamma_p$, both inside the 
molecular clouds and in the region between the clouds. Evidently, for 
proton energies $E_p \le 10^{18}$ eV, the energy loss rate is dominated 
by $pp$ scatterings. For simplicity, we will therefore also ignore the 
energy-loss rates due to synchrotron radiation, and Compton scattering
in calculating the particle trajectories.

\section{Proton Propagation}
There are two basic approaches to simulating the propagation of protons
through the interstellar medium. These may be summarized as follows:
\begin{enumerate}
\item solving the Lorentz force equation, ${\bf F} = q {\bf v} \times {\bf B}$, 
for each individual particle, or
\item solving the diffusion equation,
\begin{equation}
\frac{\partial W}{\partial t} + \frac{\partial}{\partial x_i}(Wv_i) - 
\frac{\partial}{\partial x_i}(\kappa_{ij} \frac{\partial W}{\partial x_j})
\nonumber
\;,
\end{equation}
for the distribution function $W(x_i,t)$, in terms of the diffusion
coefficients $\kappa_{ij}$.
\end{enumerate}
Calculating the positions of a large number of protons using the Lorentz 
force equation is very time-consuming. On the other hand, we only have 
rough estimates of the diffusion coefficients needed to solve the diffusion 
equation. For this study, we have devised a hybrid approach in which
we calculate these coefficients using the motion of individual particles
traced with the Lorentz equation. Our method is a two-step approach in 
which we first follow the trajectories of a select number of individual 
protons, and then use the statistics from this short simulation to 
model their motion over a much longer period of time.  In this way, we 
make no assumption about the protons' diffusion and the time needed to 
run the simulation is significantly reduced.

\subsection{Individual Proton Trajectories}
We solve the Lorentz force equation for 1,000 protons with kinetic energy 
$T_p$ moving in a uniform medium with conditions representative of a
a giant molecular cloud interior, and then of the region between the 
clouds. The protons diffuse away from their point of origin by 
random-walking through the turbulent magnetic field, generated at each 
point along their path using the method described in section 3. Unlike 
the application of Giacalone and Jokipii (1994), however, we use 200 
values of the wavenumber $k$ evenly spaced on a logarithmic scale between 
$k_{min} = {2 \pi}/{10 r_{gyr}}$ and $k_{max} = {2 \pi}/ {0.1 
r_{gyr}}$, in terms of the gyroradius $r_{gyr} = {\gamma_p m_p v}/{e B}$ 
of a proton with Lorentz factor $\gamma_p$.

For each set of pre-selected environmental conditions, the total time 
$\tau$ we follow each proton is chosen small enough that it loses
no more than $1\%$ of its energy (${\Delta E}/{E} < 0.01$), yet large 
enough that it will gyrate many times. This time interval is further
sub-divided into much smaller time steps $dt = \chi\,{2\pi \gamma_p 
m_p}/{e B}$, where $\chi$ is a random number such that $0 \leq \chi \leq 1$.

To calculate the proton's position after each $dt$, we first rotate to a primed 
set of coordinates in which the magnetic field points along the $z'$-direction 
(${\bf B'} = B_0 \hat{\bf k'}$), solve the Lorentz force equation, and then rotate 
back to the original, unprimed coordinate system.  This process is repeated at each 
step of the random-walk until the total time limit has been reached.  

\begin{figure}
\center{\includegraphics[width=10cm]{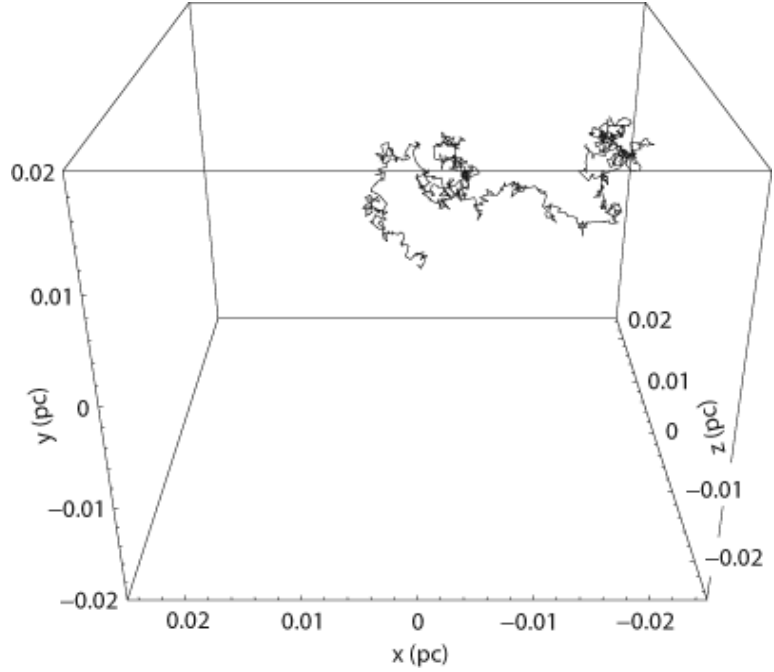}
\caption{Trajectory of a 1-TeV proton through the inter-molecular cloud region.  
The proton travels for a total time of $3.33 \times 10^8$ seconds and
covers a net distance of 0.02 parsecs. Every 50$^{th}$ step is shown.}
\label{fig:sampletrack}}
\end{figure}

\subsection{Proton-Propagation Statistics}
The data for the individual protons whose trajectories have been calculated
in the manner described above are plotted in Figure~5, showing their
occupation as a function of distance from their point of origin. The
distribution is strongly dependent on the proton energy, so this process
must be repeated for a wide range of proton kinetic energies $T_p$.
The distribution can be modeled adequately with a Gaussian function
\begin{equation}
N(r) = N_0 ~ e^{{-(r - \bar{r})^2}/{2 \sigma^2}}\;,
\label{eq:Gaussian}
\end{equation}
where the average distance $\bar{r}$ and standard deviation $\sigma$ are to be 
determined using a $\chi^2$ fitting of the data with the Levenberg-Marquardt 
method. 

\begin{figure}
\center{\includegraphics[width=10cm]{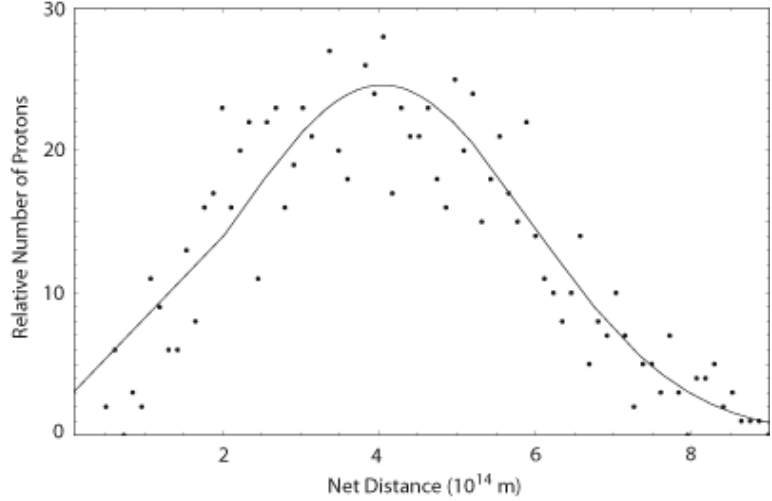}
\caption{This figure shows the distribution of 1000 1-TeV protons as a function of 
position after $3.33 \times 10^8$ seconds. The dots are data from the short
simulation, the line is the Gaussian fit to the data with $\bar{r} = 4.05\times10^{14}$ 
meters and $\sigma = 1.91\times10^{14}$ meters.}
\label{fig:Gaussian}}
\end{figure}

We repeat this procedure over the energy range $10^{11}$ eV $\leq T_p < 
T_{p, trans}$, where $T_{p, trans}$ is the energy at which the protons 
transition from diffusive random-walk behavior to (effectively) straight-line 
motion. From a practical standpoint, this transition may be taken to occur when 
the total distance traveled by a proton becomes less than $10 r_{gyr}$.  
The time elapsed during the simulation (see Table 2) is fixed by the requirement that
protons lose a given fraction of their initial kinetic energy. In our calculations,
these values are $\Delta T_p/T_p = 0.01$ for environmental conditions matching those 
found within the molecular clouds, and $\Delta T_p/T_p = 0.001$ in the inter-cloud 
medium. We have found the following formulations of the average net distance traveled 
and standard deviation in terms of the proton kinetic energy to be useful 
representations of the data:
\begin{subequations}
  \begin{equation}
    \ln({\bar{r}/1\;{\rm m}}) = a_r + b_r ~\ln(T_p/1\;{\rm eV})
  \label{subeq:r}
  \end{equation}
  \begin{equation}
    \ln(\sigma/1\;{\rm m}) = a_{\sigma} + b_{\sigma} ~\log(T_p/1\;{\rm eV})\;,
  \label{subeq:sigma}
  \end{equation}
\end{subequations}
where $a_r$, $b_r$, $a_{\sigma}$, and $b_{\sigma}$ are constants determined from 
$\chi^2$ optimization. A sample of these constants is given in Table 2 for three 
different magnetic field intensities. 

\begin{table}
\caption{Fitting Coefficients as a Function of Magnetic Field Strength, Ambient 
Proton Density, and Elapsed Time}
\footnotesize
\begin{tabular}{lllllll}
 $B\; (\mu$G)\qquad\qquad&  $n_p\; ($cm$^{-3})$&  $t\; (s)$&  
$a_r$\qquad\qquad\qquad&  $b_r$\qquad\qquad\qquad&
 $a_{\sigma}$\qquad\qquad\qquad&  $b_{\sigma}$\qquad\qquad\qquad \\
\hline
 10&  100&  $1.67 \times 10^{10}$&  9.98033&  0.499622&  9.68435&  0.497083 \\
 100& 100&  $1.67 \times 10^{10}$&  9.50314&  0.497989&  9.16946&  0.498592 \\
 1000&  10$^4$&  $1.67 \times 10^{9}$&  8.48292&  0.499238&  8.13816&  0.500453 \\
\end{tabular}
\end{table}

\subsection{The Proton Distribution}
The use of equations~(\ref{eq:Gaussian}), (\ref{subeq:r}), and (\ref{subeq:sigma})
permits us to propagate the protons with any given energy $T_p$ quickly and
accurately. If $T_p \geq T_{p, trans}$, the proton moves along
a straight line to its next interaction point (usually a $pp$ scattering event). 
If $T_p < T_{p, trans}$, the motion is diffusive and Monte-Carlo methods are 
employed with the use of equations~(\ref{eq:Gaussian}), (\ref{subeq:r}), and 
(\ref{subeq:sigma}). 

\section{The Particle Cascade}
As the protons diffuse, a fraction of them undergo a $pp$ scattering event
after traveling a total path length $dx$, according to the rate
\begin{equation}
 \frac{dN}{dx} = -N n_H \sigma_{pp}.
\label{eq:dNdx}
\end{equation}
These $pp$ scatterings create pions that decay, ultimately producing 
gamma rays, electrons, positrons, and a host of neutrinos. A simulated
image of the gamma ray emission may be obtained by calculating and mapping 
the spectra of the secondary particles produced in this cascade.

Inelastic scatterings between high-energy protons $n_p(E_p)$ and ambient 
protons $n_H$ result in a $\pi^0$ emissivity 
\begin{equation}
 Q^{pp}_{\pi_0} = c n_H \int_{E_{th}(E_{\pi^0})} dE_p n_p(E_p) 
\frac{d\sigma (E_{\pi^0}, E_p)}{dE_{\pi^0}}\;,
\label{eq:pionemissivity}
\end{equation}
where $E_{th}(E_{\pi^0})$ is the minimum proton energy needed to 
produce a pion with energy $E_{\pi^0}$.  The neutral pion decay $\pi^0
~ \rightarrow ~2 \gamma$ leads to a gamma-ray emissivity
\begin{equation}
 Q_{\gamma}(E_{\gamma}) = 2 \int_{E^{min}_{\pi^0}(E_{\gamma})} dE_{\pi^0} 
\frac{Q^{pp}_{\pi^0}}{(E^2_{\pi^0} - m^2_{\pi^0} c^4)^{1/2}}\;,
\label{eq:gammaemissivity}
\end{equation}
where $E^{min}_{\pi^0}(E_{\gamma}) = E_{\gamma} + m^2_{\pi^0} c^4/(4 E_{\gamma})$.

For protons with energies less than 3 GeV, the differential $\pi^0$ cross 
section may be determined using the isobar model of Stecker (1970) (see also 
Dermer 1986).  The expressions needed to calculate the low-energy cross 
section are quite detailed and lengthy so they are not repeated here, but may 
be found in, e.g., the Appendix of Fatuzzo and Melia (2003). Above 7 GeV, 
we use the scaling approximation of Blasi and Colafrancesco (1999),
\begin{equation}
 \frac{d\sigma(E_p,E_{\pi^0})}{dE_{\pi^0}} = \frac{\sigma_0}{E_{\pi^0}}f_{\pi^0}(x)
\,,
\end{equation}
where $x$ = $E_{\pi^0}/E_p$, $\sigma_0$ = 32 mbarn, and 
\begin{equation}
 f_{\pi^0}(x) = 0.67(1 - x)^{3.5} + 0.5 e^{-18x}
\end{equation}
takes into account the energy-dependent pion multiplicities that occur at 
high energies. In the intermediate (3-7 GeV) range, a linear combination 
of the low- and high-energy forms is used.

\section{Results and Discussion}
Our first attempt to match HESS's map of the diffuse emission is
based on the assumption that all of the cosmic-ray protons originate 
from Sagittarius A*, since only $\sim 1/3$ of the hadrons required to 
account for the central TeV source HESS~J1745-290 actually interact 
with the circumnuclear disk surrounding the black hole. The remaining
proton efflux could in principle fill the interstellar medium out to 
$|l|\sim 1^\circ$, and account for the diffuse TeV glow associated 
with the molecular gas dispersed along the galactic ridge (see 
Figures~1 and 2).

\begin{figure}
\center{\includegraphics[width=14cm]{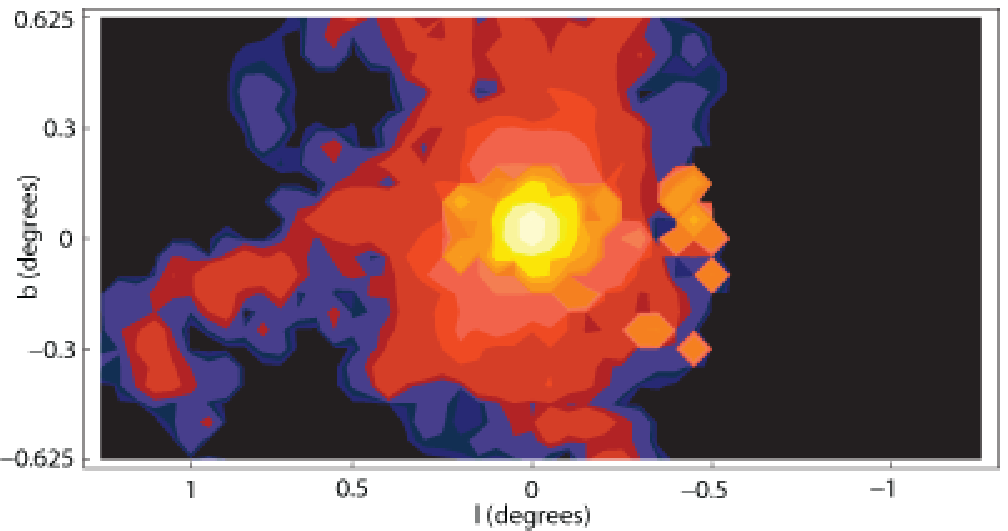}
\vskip 0.1in
\includegraphics[width=14cm]{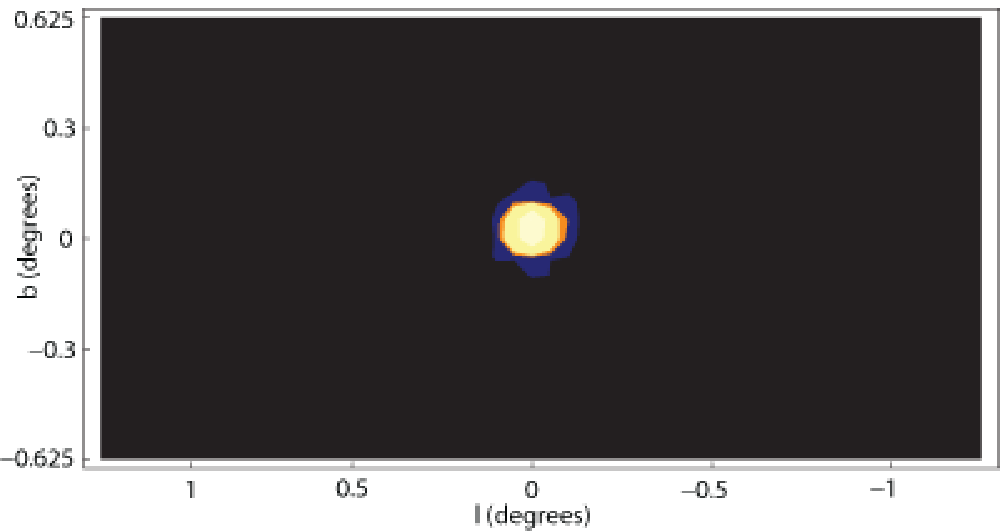}
\caption{Gamma ray intensity in the energy range 0.2--10 TeV, calculated 
from equations~(13) and (14), assuming Sagittarius A* as the sole source 
of relativistic protons, and an average magnetic field of $10\mu$G in 
the inter-cloud medium. Top panel: total intensity map spanning 8 orders
of magnitude from the brightest regions (white, near the centre) to the 
lowest (blue, near the edge). Bottom panel: intensity map spanning the
highest intensity range, decreasing by a factor two from the centre
(white) to the edge (blue).}
\label{fig:resultone}}
\end{figure}

The gamma-ray count map resulting from our simulation with Sagittarius A* 
as the sole source of high-energy protons, calculated from equations~(13)
and (14), is shown in Figure~\ref{fig:resultone}. The top panel is a plot 
of the full range of photon counts spanning eight orders of magnitude in
intensity. However, the actual HESS diffuse TeV emission map (Aharonian 
et al. 2006) covers a much smaller dynamic range, dropping by only a factor of 
two from the highest intensity pixels down to the lowest.  Thus, in order
to make a direct comparison with the HESS data, we also show in the lower panel
the corresponding calculated intensity map with the same factor-2 intensity
range. In either case, it is clear that the protons from Sagittarius A* acting 
as the lone source of cosmic rays cannot explain the observed diffuse gamma-ray 
emission. The very evident peak in the TeV emission associated with the
black hole itself may still be a viable explanation for the central source
HESS~J1745-290, but subtracting this from the rest of the diffuse emission
produces a TeV map centred on Sagittarius A*, extending out only a fraction 
of a degree, or at most only about $1^\circ$ if we include the full 8 orders 
of magnitude in intensity. 

\begin{figure}
\center{\includegraphics[width=14cm]{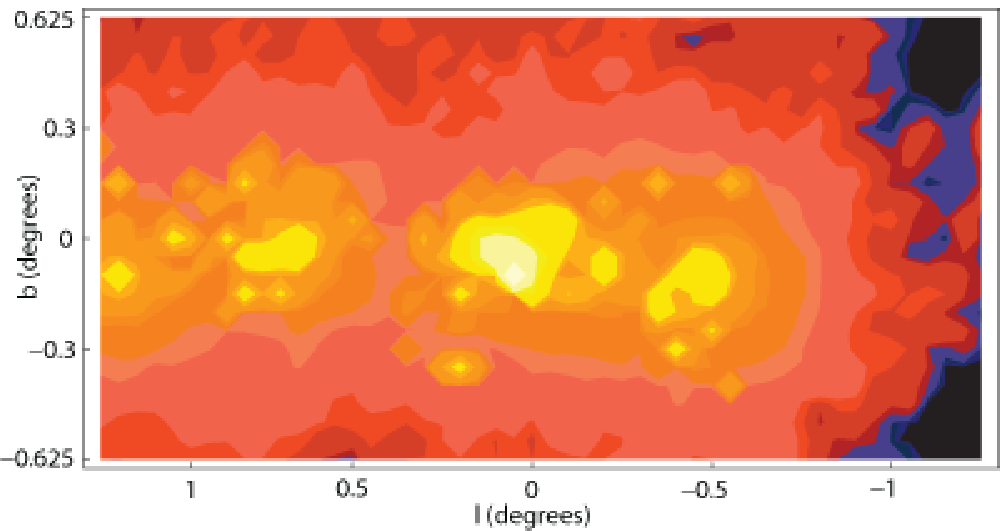}
\vskip 0.1in
\includegraphics[width=14cm]{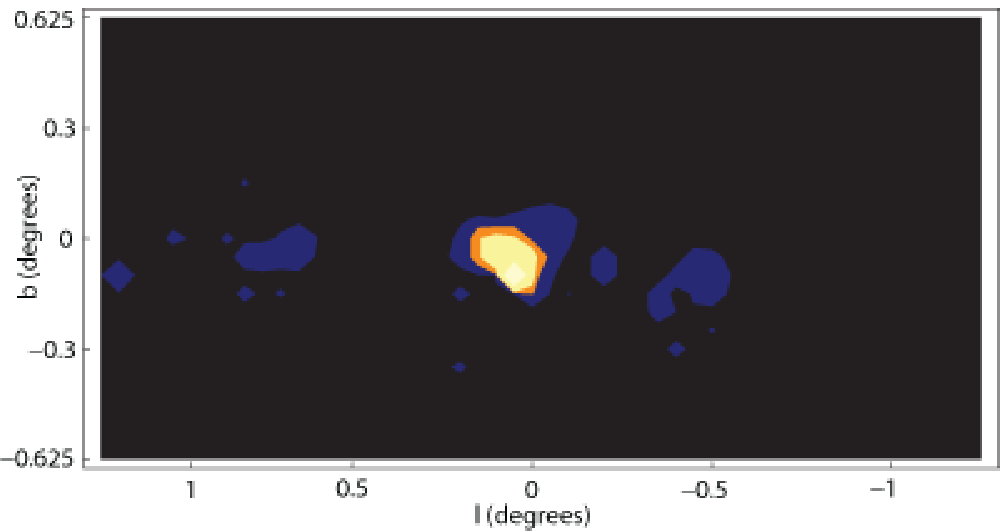}
\caption{Same as figure~6, except here for protons injected into the inter-cloud
medium by 5 distinct sources. The diffuse emission is more pronounced than that
in figure~6, though the emission tends to be concentrated on the point sources,
and is only weakly correlated with the molecular gas distribution shown in 
figures~1 and 2.}
\label{fig:resulttwo}}
\end{figure}

Our conclusion is that as long as the magnetic field surrounding Sagittarius
A* is at least partially turbulent (we here assumed that half of the magnetic
energy is turbulent), then the hadronic diffusion away from the galactic centre 
is too slow to account for $pp$ scattering events with molecular gas more than 
a fraction of a degree (i.e., only tens of parsecs) out along the ridge. 

We next considered a situation in which the cosmic-ray protons originate from
several point sources distributed along the galactic plane. In addition to
diffuse emission, HESS has discovered several points sources in this
region, including HESS~J1745-290, typically associated with supernova 
remnants or pulsar wind nebulae. The HESS map reveals five TeV ``hot spots" 
in the region $|l| \le 1.25^\circ$ and $|b| \le 0.5^\circ$. In the second
simulation, we therefore distributed the proton injection among 5 individual
point sources, assigning them the observed latitudes and longitudes, and
a random $z$ (line-of-sight) coordinate consistent with positions between 
the molecular clouds.  

\begin{figure}
\center{\includegraphics[width=14cm]{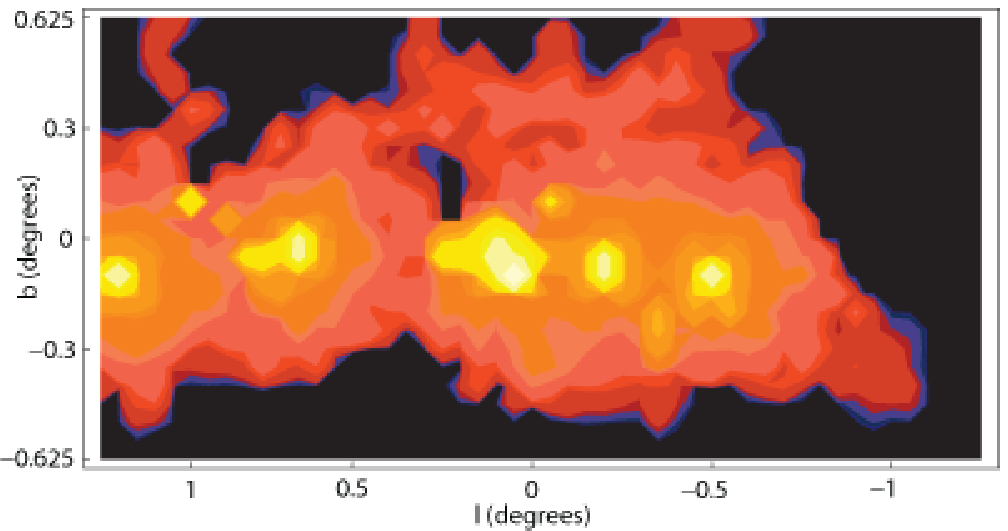}
\vskip 0.1in
\includegraphics[width=14cm]{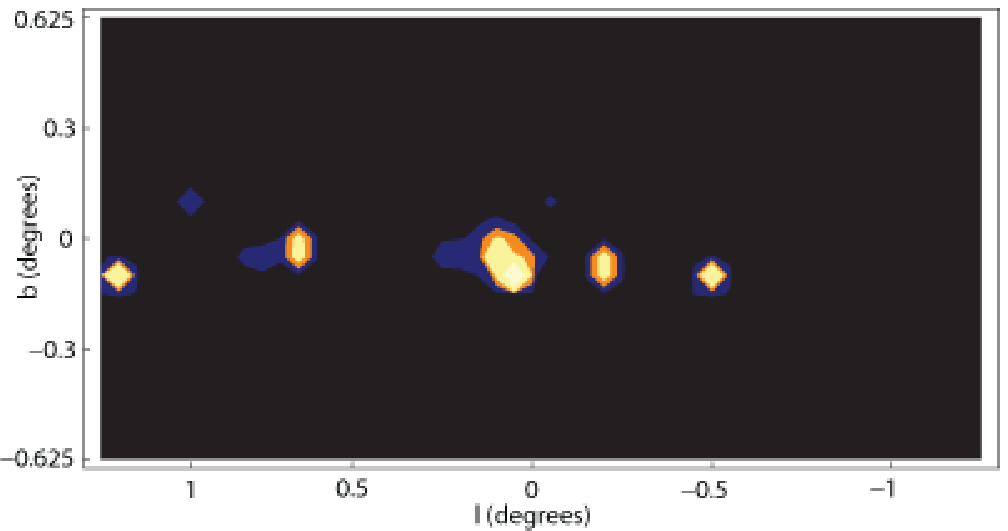}
\caption{Same as figure~7, except here for a magnetic field of $100\;\mu$G.
The stronger magnetic field confines the protons more strongly, enhancing
the point-like emissivity at the expense of the diffuse flux. Again, the
correlation with the molecular gas distribution (figures~1 and 2) is very weak.}
\label{fig:resultthree}}
\end{figure}

To further test the dependence of our proton diffusion (and the associated
synthetic TeV map) on the magnetic field intensity, we simulated the hadronic
propagation for two values of $B$, $10\;\mu$G (Figure ~\ref{fig:resulttwo}) 
and $100\;\mu$G (Figure~\ref{fig:resultthree}). As expected, increasing the 
strength of the magnetic field confines the protons more strongly, and 
therefore restricts the TeV emission to more compact regions surrounding
the sources. Here too, a dominant feature of the gamma-ray maps is 
the evident concentration of TeV emission near the injection sites, which
might account very well for the point sources themselves. But the diffusion
scale is still too small, for both values of magnetic field, to permit
the protons to fill the medium between the clouds with a sufficiently
dispersed cosmic-ray population to account for the HESS intensity map.
Decreasing the magnetic field further, to a value $\sim 1\,\mu$G
helps, but the synthetic map always contains intensity gradients revealing
the location of the point sources, in contrast with the actual TeV map,
which shows a strong correlation of the gamma-ray emissivity only with the
measured concentration of molecular gas and none of the individual point
sources.

It is quite evident from these two sets of simulations that if the TeV 
photons are to be produced in hadronic interactions involving the molecular 
gas, the relativistic protons themselves must be accelerated in situ, 
throughout the interstellar medium. The fact that their diffusion length 
is at most only a fraction of a degree, means that even a remote acceleration 
site, well away from the molecular clouds, cannot produce the required cosmic-ray
distribution. 

\begin{figure}
\center{\includegraphics[width=14cm]{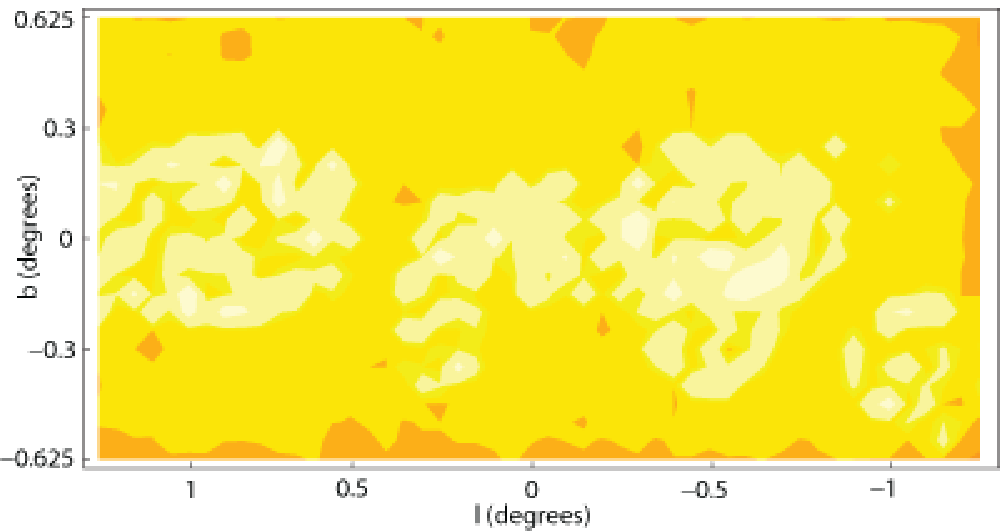}
\vskip 0.1in
\includegraphics[width=14cm]{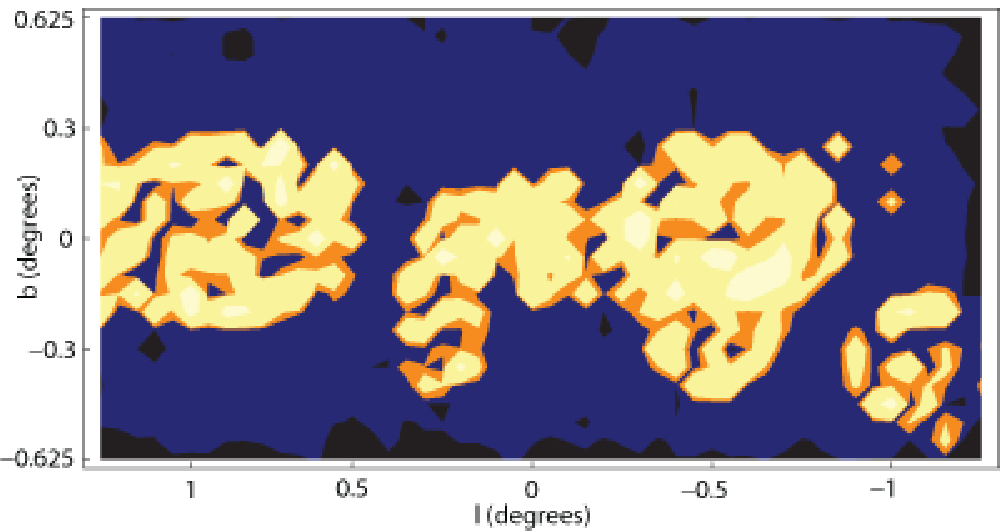}
\caption{Unlike the simulations displayed in figures~6, 7, and 8, the
gamma-ray intensity map shown here is produced entirely by relativistic
protons injected throughout the inter-cloud medium (e.g., by 
second-order Fermi acceleration). The magnetic field is assumed to have
an average value of $10\;\mu$G. The correlation between the gamma-ray
emissivity and the molecular gas distribution (figures~1 and 2) is quite
evident. However, the protons responsible for producing this diffuse TeV 
emission do not have the same distribution as the cosmic-ray population
measured at Earth. These hadrons therefore apparently represent a distinct
population at the galactic centre.}
\label{fig:resultfour}}
\end{figure}

To examine whether protons accelerated throughout the inter-cloud medium can 
in fact produce the observed diffuse TeV emission, we therefore also simulated 
a situation in which the protons emerge uniformly via, e.g., second-order 
Fermi acceleration off the turbulent magnetic field. In this calculation,
the starting point for each proton falls somewhere between the molecular 
clouds but is otherwise chosen randomly. Figure~\ref{fig:resultfour} shows 
the synthetic TeV intensity map resulting from this model, demonstrating a 
strong correlation with the location (in projection) of the molecular gas
(see figures~1 and 2), in excellent agreement with the HESS observations. 

\section{Conclusion}
In summary, then, we have found that the conditions at the galactic
centre preclude a point-source origin for the cosmic rays responsible 
for producing the diffuse TeV emission correlated with the molecular
gas distributed along the ridge. The supermassive black hole Sagittarius
A* may be associated with the central object HESS~J1745-290, but its
hadronic efflux cannot extend out to $\sim 1^\circ$ since the protons
lose their energy or scatter with the ambient medium on much smaller
scales. Distributed point sources offer the possibility of extending
the diffuse TeV emission to greater distances from the centre, but
they too would produce a morphology with centrally peaked emission
regions not consistent with the HESS map. Only cosmic rays accelerated
throughout the inter-cloud medium can produce a diffuse TeV glow
consistent with the observations.

An important question is therefore whether the particle acceleration
occurs within the interstellar medium, or whether the cosmic rays
emerge from any known population of sources distributed throughout
the galactic ridge. From our simulations, we infer that the gamma 
ray emissivity associated with any given object drops by a factor
$\sim 2$ within a distance of roughly $0.1^\circ$. Since this is
effectively the contour range of the HESS maps, individual sources
would not be evident as long as their angular separation were less than 
this value. In a projected area $\sim 2^\circ\times 1^\circ$,
this would require about 50 individual sources. But the total number 
of TeV sources detected by HESS (many of them presumably pulsar
wind nebulae) was far smaller than this. In addition, only $\sim 5$ 
low-mass X-ray binaries have been identified in this region 
(Bird et al. 2007), and no other class of object with a volume
density greater than this is known to be a strong source of relativistic
hadrons. It is therefore likely that the particles are not being
injected into the interstellar medium by individual objects.

Since the relativistic protons pervading the inter-cloud medium at the
galactic centre are accelerated throughout the plane, second-order 
Fermi acceleration may be dynamically important in this region. 
A reasonable extension of this work will therefore be the inclusion 
of stochastic particle acceleration by the same turbulent magnetic 
field responsible for the particle diffusion. 

Future extensions of this model should include a self-consistent
calculation of the longer wavelength emission produced by particles
in the $pp$-induced cascade. At the very least, the synchrotron
emission due to the secondary leptons spiraling around the
assumed magnetic field must not exceed the diffuse radio glow 
along the ridge. In principle, the TeV and radio intensity maps, 
used together, might produce a unique measurement of the magnetic 
field under the assumption that a single hadronic process is responsible 
for both spectral components. The results of this calculation will be 
reported elsewhere.

\vspace*{-0.3cm}
\section*{Acknowledgments}

This research was partially supported by NSF grant 0402502 at the
University of Arizona, and a Miegunyah Fellowship at the University
of Melbourne.

\end{document}